\documentclass[a4paper]{article}
\usepackage{graphicx,epsfig}
\pdfoutput=1
\DeclareGraphicsExtensions{. jpg,. pdf, . mps, .png, .eps, . ps, . EPS}
\usepackage{color}
\usepackage[english]{babel}
\usepackage{amsmath}
\usepackage {amsfonts}
\usepackage {amssymb}
\usepackage {graphicx}
\usepackage {fancyhdr}
\usepackage{newlfont}
\usepackage{epigraph}
\usepackage{indentfirst}
\usepackage[table]{xcolor}
\usepackage{booktabs}
\usepackage{authblk}
\usepackage{verbatim}
\usepackage{colortbl}
\usepackage{subfigure}
\usepackage[T1]{fontenc}
\usepackage[utf8]{inputenc}
\usepackage{url}

\usepackage{etoolbox}
\let\bbordermatrix\bordermatrix
\patchcmd{\bbordermatrix}{8.75}{4.75}{}{}
\patchcmd{\bbordermatrix}{\left(}{\left[}{}{}
\patchcmd{\bbordermatrix}{\right)}{\right]}{}{}

\date{}

\title{Characterization of DNA methylation as a function of biological complexity via dinucleotide inter-distances}
\author[1]{Giulia~Paci}
\author[3]{Giampaolo~Cristadoro}
\author[4]{Barbara~Monti}
\author[2,3]{Marco~Lenci}
\author[3]{Mirko~Degli~Esposti}
\author[1,2]{Gastone~Castellani}
\author[1,2]{Daniel~Remondini\thanks{Corresponding author: daniel.remondini@unibo.it}}

\affil[1]{Department of Physics and Astronomy, University of Bologna, Viale~B.~Pichat~6/2, 40127~Bologna, Italy}
\affil[2]{INFN, Bologna Unit, Viale~B.~Pichat~6/2, 40127~Bologna, Italy}
\affil[3]{Department of Mathematics, University of Bologna, Piazza~di~Porta~S.~Donato~5, 40126~Bologna, Italy}
\affil[4]{Department of Pharmacy and Biotechnology, University of Bologna, 
Via~S.~Donato~15, 40127~Bologna, Italy}

\begin{document}
\newcommand{\beq}{\begin{equation}}
\newcommand{\eeq}{\end{equation}}
\newcommand{\bea}{\begin{eqnarray}}
\newcommand{\eea}{\end{eqnarray}}
\newcommand{\gt}{\tilde{g}}
\newcommand{\mt}{\tilde{\mu}}
\newcommand{\et}{\tilde{\varepsilon}}
\newcommand{\ct}{\tilde{C}}
\newcommand{\bt}{\tilde{\beta}}

\newcommand{\avg}[1]{\langle{#1}\rangle}
\newcommand{\Avg}[1]{\left\langle{#1}\right\rangle}
\newcommand{\cor}[1]{\textcolor{red}{#1}}

\maketitle

\begin{abstract}
We perform a statistical study of the distances between successive occurrencies of a given dinucleotide in the DNA sequence for a number of organisms of different complexity.
Our analysis highlights peculiar features of the dinucleotide CG distribution in mammalian DNA, pointing towards a connection with the role of such dinucleotide in DNA methylation.  
While the CG distributions of mammals exhibit exponential tails with comparable parameters, the picture for the other organisms studied (e.g., fish, insects, bacteria and viruses) is more heterogeneous, possibly because in these organisms DNA methylation has different functional roles.
Our analysis suggests that the distribution of the distances between dinucleotides CG provides useful insights in characterizing and classifying organisms in terms of methylation functionalities.
\end{abstract}

\maketitle


\section{Introduction}

The statistical analysis of DNA coding and non-coding sequences has revealed structures and correlations that go beyond the extent of short-range models, for example uncovering scale-invariant properties of the sequence as a whole. Results in this direction date back to the early 80's  \cite{Trifonov85, Peng1992, Voss1992, Li1994, Peng1995, Mantegna95, Arneodo1995, Allegrini1996, Voss1996}. 
Despite these early observations, the functional role and nature (if any) of such long-range correlations are still to be clarified \cite {Li1997, Herzel1997, Audit2001, Trifonov11, Arneodo2011}. 
Other statistical measures on genetic sequences have also been investigated,  like entropies \cite{Loewenstern1999,Schmitt1997,Grosse2000} or return time statistics  for  specific oligonucleotides \cite{Afreixo11,Afreixo09}. 
In particular,  the first-return-time distribution  proved to be  a powerful tool to investigate the statistical properties of symbolic sequences in general. 
The idea dates back to the pioneering work of Poincar\'e \cite{Poincare890}, on  the trajectories of bounded dynamical systems. 
Thereafter, much effort has been devoted to the analysis of symbolic trajectories in very different contexts:  stochastic processes \cite{Durrett10}, biological data \cite{Pennetta12} and literary texts \cite{Altmann12}, to mention just a few. 
These tools have been applied to genomics in different forms for about the last 10 years \cite{Rossi04, Nair05}, becoming nowadays rather common, e.g. for the reconstruction of the phylogenetic tree \cite{Afreixo09, Afreixo11}, for the detection of CpG islands \cite{Afreixo15} and for the characterization of long-range correlations in DNA  \cite{Frahm2012}.

While the identification and quantification of  the statistical features of a genetic sequence can be instrumental in understanding certain properties of its primary structure, a more biologically motivated study can bring new insight on these  patterns.
The statistical analysis of DNA sequences can reveal functional and structural properties of biological relevance, and also, in principle,  universal features that go beyond the single organism, which can help  characterize and classify different levels of organism complexity.
It can also help comprehend complex mechanisms such as chromatin structure \cite{Choy10,Guelen08} and epigenetic regulation \cite{Yuan12,Pellegrini10}. 
Moreover, it is nowadays recognized that in complex organisms the non-coding regions of DNA, the once-called "junk DNA", are continuously annotated with novel regulatory functions \cite{Esteller11,Rinn12}. 
A deeper knowledge of the characteristics of DNA structure (not only the coding part) could help understand the effects of some pathologies that involve mutated genes with a structural rather that a functional role, as in the case of laminopathies \cite{Misteli06,Dekker13,Esteller13}.

\section{Methods}

In this study we employ a distance-based approach to characterize the distribution of dinucleotides inside human and other genomes. 
In particular, for a  given  sequence and  a given dinucleotide, we compute the distance (counted in number of bases) between two consecutive occurrences of that dinucleotide.  

More precisely, consider a  sequence $s=\{ s_j \}_{j=1}^{N}$ where $s_j$  take value in the alphabet  $\{A,C,G,T \}$. For a given dinucleotide $XY$ with  $X,Y \in \{A,C,G,T \}$, construct the sequence of  indexes where XY occurs:   $\{ r_j | s_{r_j}s_{r_{j}+1} = XY \}$. The sequence of inter-distances $\{ \tau_j \}$  is then computed from the difference of successive indexes $\tau_j=r _{j+1}-r_j$.  This corresponds to the choice of an overlapping-window frame\footnote{For  dinucleotides  of the type $XX$ ($X \in \{A,C,G,T \}$), we removed overlapping occurrences, namely, the case of the subsequence  $XXXX$ is considered  as two $XX$ dinucleotides with a distance of two, and all distances are subtracted by one at the end of the process in order to obtain a minimum distance of 1. }.  Here we focus on the  relative frequencies of  such distances, that is: 
\begin{equation}
p(\tau):= \frac{ \#\{ j | \tau_j =\tau \} }{ \#\{ \tau_j \} }
\end{equation}

For all the  considered organisms, the different sequences of inter-dinucleotide distances $\{\tau_j \}$,  for  each of the 16  dinucleotides, were computed from the whole genome sequence obtained by concatenating all chromosomes\footnote{The sex chromosomes have been excluded from the analysis.}. 
Note that, since  we focus on the probability distribution $p(\tau)$  only, the precise order of the concatenation has a negligible effect on the results. 
Finally, unknown nucleotides (corresponding to the symbol $N$ on the DNA sequence) were removed from the sequences to be analyzed. 

Logarithmic and double-logarithmic plots of the distributions were used to visually inspect the exponential or power-law behavior of their tails.
Moreover, a  quantitative estimation of the differences between dinucleotide distributions was obtained using the Jensen-Shannon distance $D_{JS}$, a symmetrized version of the Kullback-Leibler divergence $D$ \cite{CoverThomas}. 
This analysis was performed between dinucleotide distributions for the same organism, and between  the distributions of the dinucleotide CG for all the studied organisms.

Given two probability distributions $P$ and $Q$, $P = p_i, i=1,...,N$, $Q = q_i, i=1,...,N$ ($\sum_i p_i = \sum_i q_i = 1$) we have:

\begin{eqnarray}
D(P|Q) = \sum_i p_i \cdot log\frac{p_i}{q_i} \\
D_{JS} = \frac{1}{2}D(P|M) + \frac{1}{2}D(Q|M) \\
M: m_i = (p_i+q_i)/2.
\end{eqnarray}

21 organisms have been analyzed for this study: \textit{Homo sapiens},  primates (\textit{Macaca mulata}, \textit{Pan troglodytes}), mammals (\textit{Bos taurus}, \textit{Canis familiaris}, \textit{Equus caballus}, \textit{Monodelphis domesticus}, \textit{Mus musculus}, \textit{Rattus norvegicus}, \textit{Ornithorynchus anatinus}), fish (\textit{Danio rerio}, \textit{Tetraodon nigroviridis}), insects (\textit{Apis mellifera}, \textit{Drosophila melanogaster}, \textit{Tribolium castaneum}), sea organisms (\textit{Ciona intestinalis}, \textit{Oikopleura diotica}), \textit{Caenorhabditis elegans}, unicellular organisms (\textit{Escherichia Coli}, \textit{Saccharomyces cerevisiae}), and a virus (human \textit{Adenovirus}). 
Most of the chosen organisms were described in two papers \cite{Sims09, Jeltsch10}. 
The details on the organisms, and the link to the DNA sequence available online, are shown in Supplementary Table 1.

Custom code was written (Python and Matlab software) to implement data import, processing and analysis.

\section{Results}

\subsection*{CG inter-distance distribution in human DNA}

The characterization of dinucleotide inter-distance distributions in human DNA reveals a striking difference of the the dinucleotide CG and all the other couples, as shown in Fig.~\ref{HomoDNA}, \ref{Sup}, and in Supplementary Figure 1.
The double logarithmic plot shows the presence of  ``heavy'' tails in the  distributions of all the remaining dinucleotides, with an algebraic decay  $p(\tau) \sim  \tau^{-b}$ , with a similar exponent close to $3$ (average exponent $b=3.3\pm 0.4$) (regression correlation coefficient $r^2\ge 0.94$ for all distributions, except for CG, see Supplementary Table 2 also for a Chi-square test comparison).

In contrast, as shown in the inset of Fig.~\ref{HomoDNA}, the tail of the CG inter-distance distribution is asymptotically exponentially decaying: $p(\tau) \sim e^{-d\cdot \tau};$ with parameter  $d=0.004\pm0.001$,   ($r^2 = 0.999$ , $\chi^2=0.044$, see Table \ref{TabFit}).

In order to find a biological meaning for such striking differences, we remark that an exponential distribution is associated with a ``characteristic  length $\lambda$ ''  between consecutive appearances of the same dinucleotide, given by the inverse of the exponential rate $d$, with a value of $\lambda\simeq 250$ bases for human DNA.
CG dinucleotides thus perform a sort of ``Bernoulli walk'' along the whole DNA sequence, at difference with the other dinucleotides for which a power-law tail implies a scale-invariant distribution. This result can be associated with the different role that CG dinucleotides have in human DNA, since they are the sites in which a methyl group can be attached by the specific enzyme family of DNA Methyltransferases \cite{Jeltsch10}.

Regarding our analyses, we remark that, since the coding regions of human DNA constitute only a small part of the overall sequence (about $1\%$ ), our statistics are mainly affected by the features of the non-coding regions, believed to have functional roles for the three-dimensional structure of the chromatin \cite{Pellegrini10} and for the regulation of transposable elements \cite{Jursch13}.
Moreover, if we consider a known structure for the dinuclotides CG  in human DNA, the so-called ``CpG islands'' \cite{Frommer87,Zhao09} (that seem to have a role in regulating the expression of the contiguous genes \cite{Razin98,Yuan12}), it is known that CG's  are at close distance between each other (inside an island).
Since we are studying the long-range interval of the inter-distance distributions (the right tail of the distributions) we can assert that our analyses are not affected by these entities in the interval considered for fitting.

\begin{figure}[htb]
\centering
\includegraphics[width=0.48\textwidth]{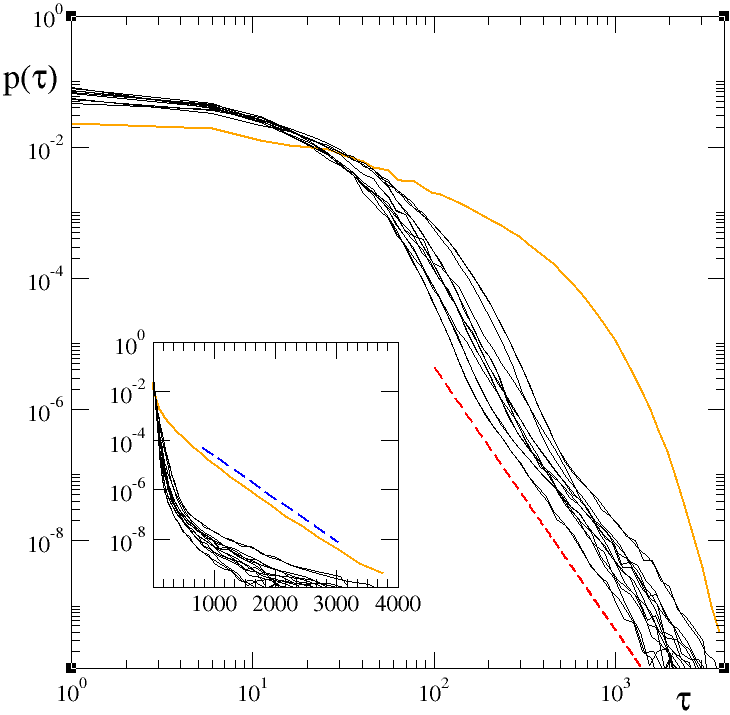} 
\includegraphics[width=0.485\textwidth]{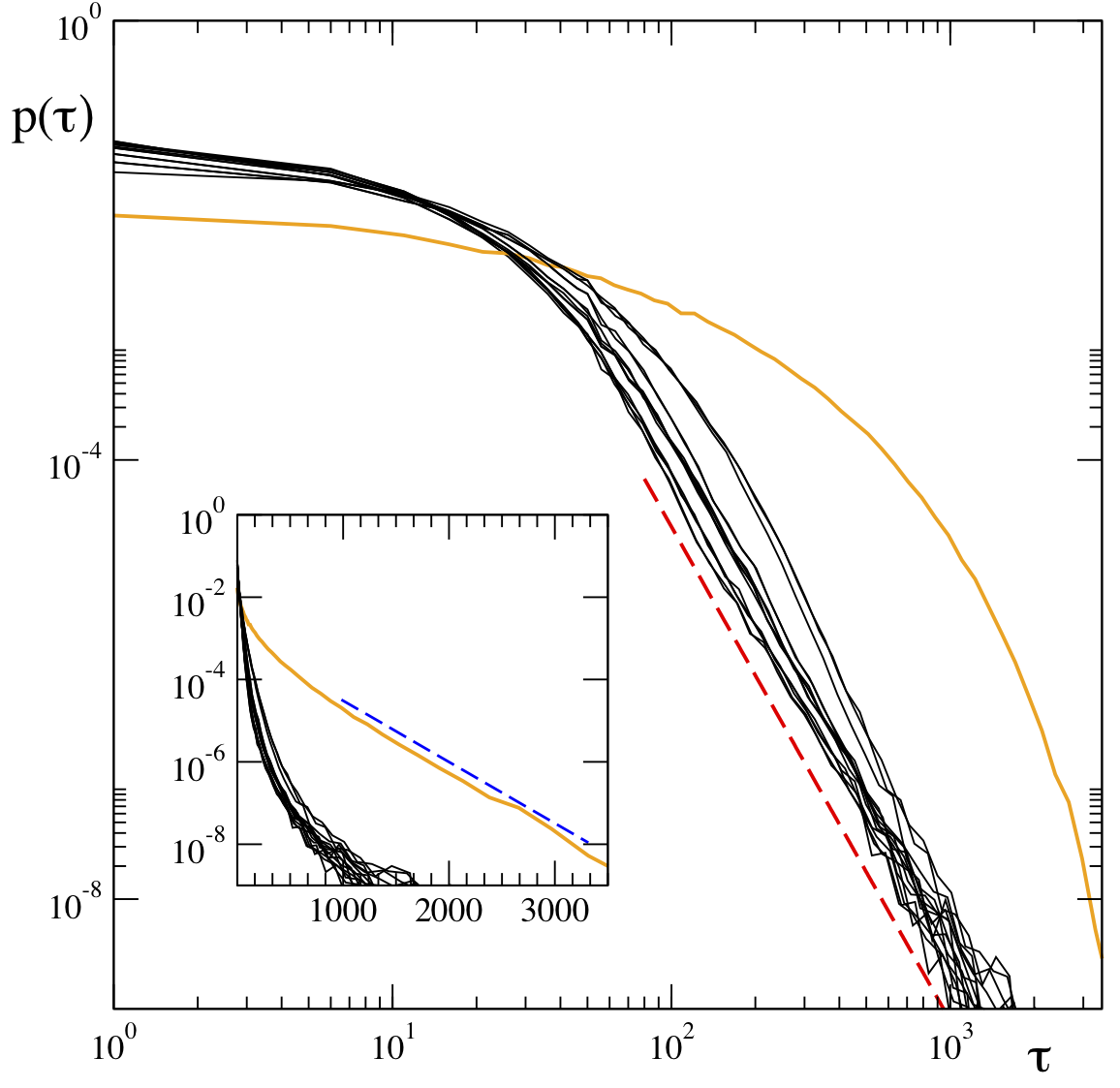} 
\caption{Plot of  the dinucleotide inter-distance distributions of \textit{Homo sapiens} (left) and \textit{Mus musculus} (right). 
In the double-logarithmic plot and in the logarithmic plot (inset) the CG distribution is colored.
Dashed lines are just a guide to the eye.}
\label{HomoDNA}
\end{figure}

\subsection*{Comparison with other organisms}

After performing our analyses on the human DNA, we have considered the DNA sequence of other organisms for comparison.
In this way we aim to find relations between the features of the CG distributions and the biological information actually available on DNA methylation mechanisms, for a large class of organisms (as described in \cite{Jeltsch10} and references therein). We remark that many results on DNA methylation in living organisms are still unknown, or at least still object of debate.

\begin{figure}[htb]
\centering
\includegraphics[width=0.9\textwidth]{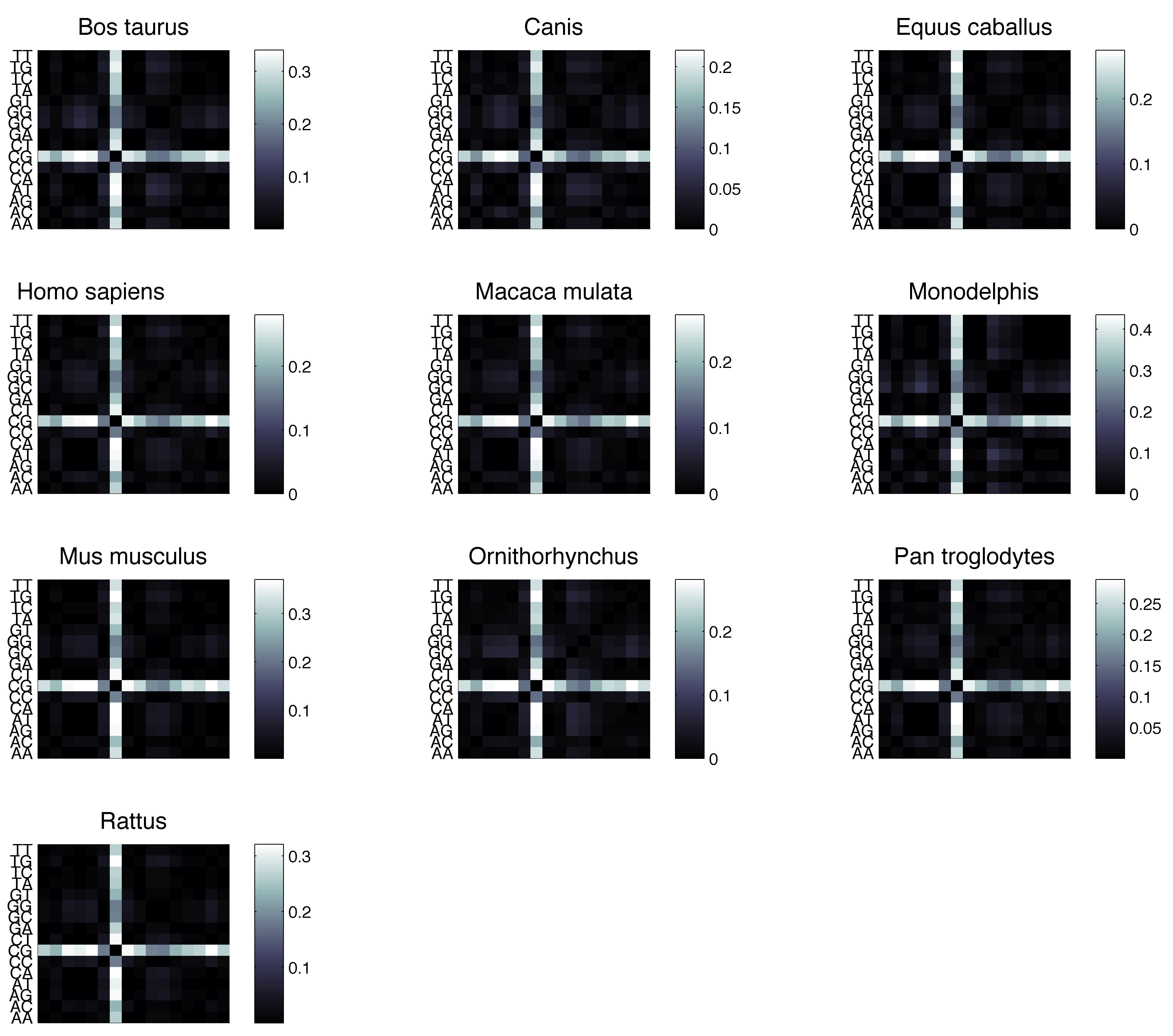} 
\caption{Plot of Jensen-Shannon distances between dinucleotide inter-distance distributions for the mammals included in the study (10 organisms).}
\label{Sup}
\end{figure}

For all the 21 organisms, we have estimated the inter-distance distributions for all the 16 dinucleotides, and compared the distributions with each other via the Jensen-Shannon distance.
We observe that a group of 10 organisms, that we can recognize as mammals, has a strikingly different distribution of CG distances, compared to the others (as shown in Fig. \ref{Sup}).
This difference is particularly evident in the tail of the CG distributions, cf. Figure \ref{HomoDNA} for man and mouse.

\begin{figure}[htb]
\centering
\includegraphics[width=0.9\textwidth]{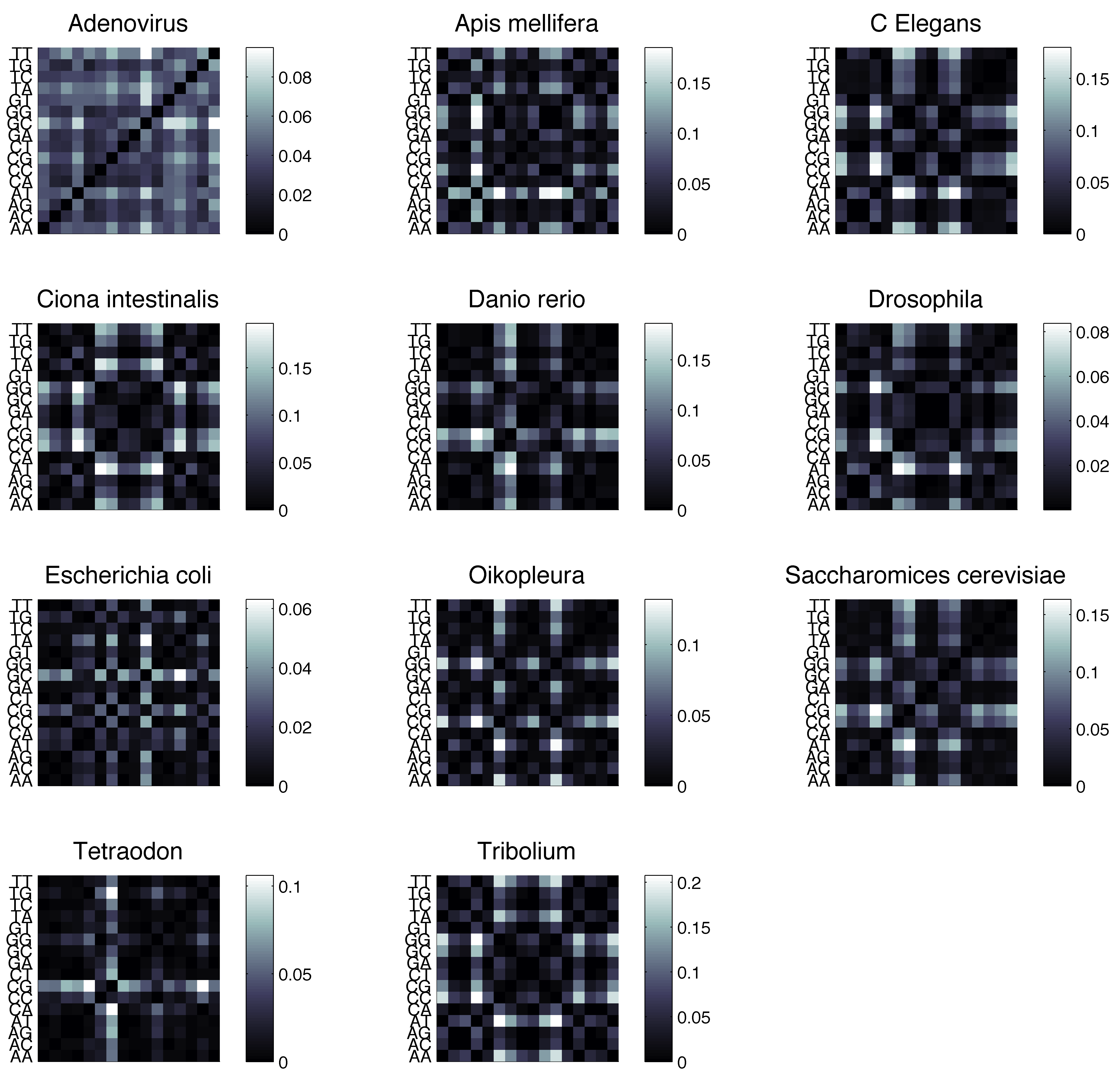} 
\caption{Plot of Jensen-Shannon distances between dinucleotide inter-distance distributions for the remaining 11 organisms included in the study.}
\label{Inf}
\end{figure}

The remaining 11 organisms show a more heterogeneous behavior, in term of the Jensen-Shannon distance: for \textit{Adenovirus, Apis}, \textit{E. Coli} and \textit{Oikopleura} no clear difference can be seen, while for the other organisms the dinucleotide CG , but also CC, GC and GG, appears different from the other distributions, showing analogous patterns of JS distances (see Fig. \ref{Inf}).

\begin{figure}[h!]
\centering
\includegraphics[width=0.65\textwidth]{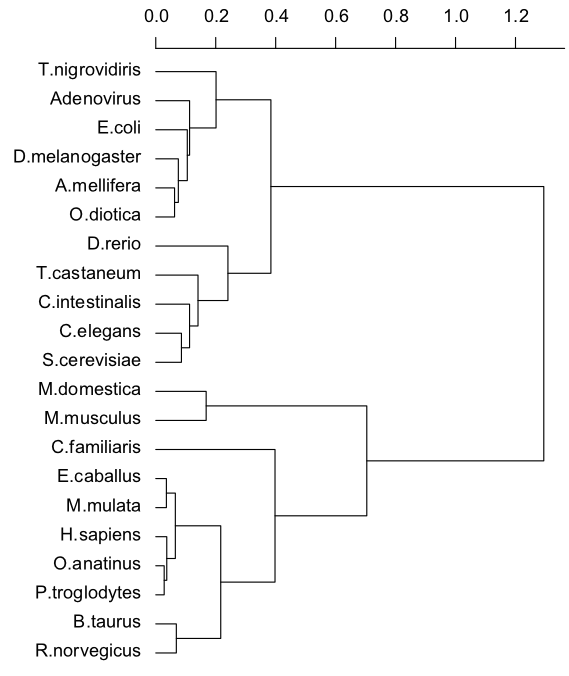} 
\caption{Dendrogram plot for all organisms, constructed using the vectors made by the JS distances between CG and the other dinucleotides. Complete linkage and Euclidean distance is used for this plot. The same two clusters are obtained by varying linkage method and distance metrics (not shown). }
\label{dendro}
\end{figure}

The  power of the JS distance in differentiating between CG and other dinucleotides can be appreciated in  Fig. \ref{dendro}: a hierarchical clustering of all organism assigns to two distinct groups mammalians and the other organisms.

\begin{figure}[htb]
\centering
\includegraphics[width=0.48\textwidth]{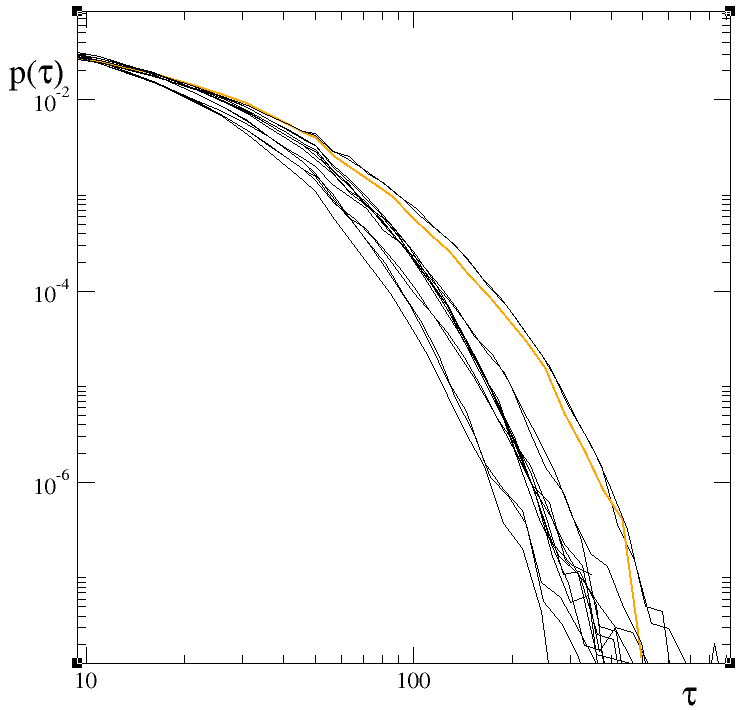} 
\includegraphics[width=0.485\textwidth]{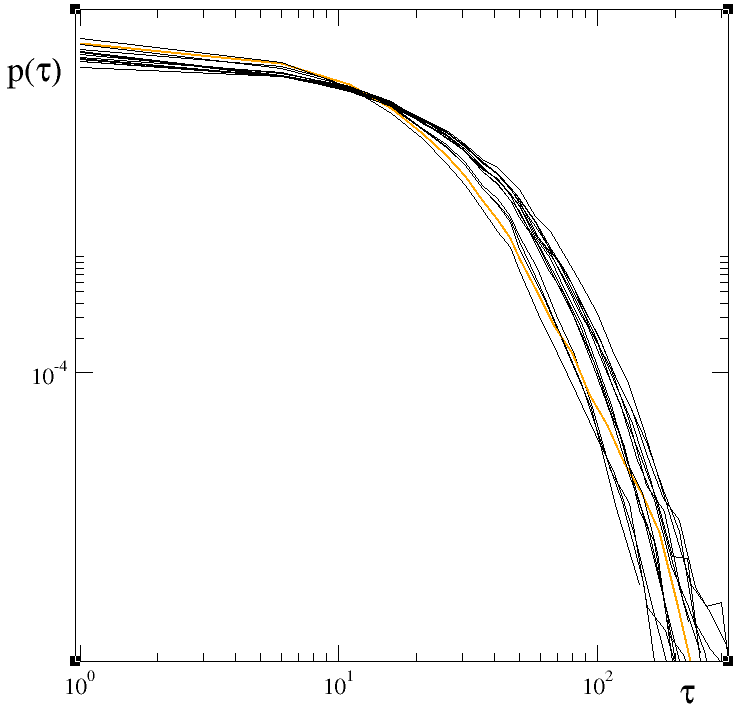} 
\caption{Double-logarithmic plot of the inter-distance distributions of \textit{D. melanogaster} (left) and \textit{E. coli} (right) dinucleotide distributions, with the CG distribution  colored differently.
}
\label{EcoliDNA}
\end{figure}

A plot of two sample organisms, \textit{Drosophila} and \textit{E. Coli}, shows that the 16 dinucleotide distributions are not as different between each other as for higher-order organisms (see Fig. \ref{EcoliDNA}).
It might be plausible that, in this group of lower-order organisms, different mechanisms related to CG methylation are present, possibly related to their different degrees of complexity, since this group comprises viruses, bacteria, a yeast strain, insects and fish.

In order to verify the goodness of this hypothesis, and also to compare the different organisms,  we have generated the cumulative distribution of the CG inter-distances for each organism, and fitted their tails  with an exponential function\footnote{We have fitted the empirical cumulative distributions in an interval of distances between 700 and 2000, to remove the effect of short distances and the possible undersampling at large distances. We have verified that reducing the lower extreme of the interval up to 300 and increasing the higher extreme up to the maximum length did not change the results significantly.}.
Since \textit{Adenoma}, \textit{E. coli} , \textit{Oikopleura} and \textit{Saccharomices} have a smaller maximum observed distance (517, 318, 679 and 308 respectively), in these cases we have fitted the whole CG distribution.

\begin{table}[!t]\small
\label{TabFit}
\centering{tb
\begin{tabular}{lccccc} 
\textbf{Organism}	& \textbf{Max}  &  \textbf{d} &  \textbf{$\lambda$} & \textbf{$r^2$} & $\chi^2$ \\ 
\hline
Bos Taurus	& 3709	& 0.0037	$\pm$	0.0001 &	272	$\pm$	1 & 0.999 & 0.032\\
Canis Familiaris	& 3248	& 0.0036	$\pm$	0.0001 &	274	$\pm$	1 & 0.999 & 0.052\\
Equus Caballus	& 2927	& 0.0047	$\pm$	0.0001 &	214	$\pm$	1 & 0.999 & 0.078\\
Homo Sapiens	& 3760	& 0.004	$\pm$	0.0001 &	252	$\pm$	1 & 0.999 & 0.044\\
Macaca Mulatta	& 3907	& 0.0042	$\pm$	0.0001 &	240	$\pm$	1 & 0.999 & 0.039\\
Monodelphis domestica	& 8123	& 0.0022	$\pm$	0.0001 &	452	$\pm$	1 & 0.999 & 0.018\\
Mus Musculus	& 4617	& 0.0034	$\pm$	0.0001 &	295	$\pm$	1 & 0.999 & 0.046\\
Ornithorhynchus anatinus	& 2841	& 0.0043	$\pm$	0.0001 &	232	$\pm$	1 & 0.999 & 0.062\\
Pan Troglodytes	& 3376	& 0.004	$\pm$	0.0001 &	248	$\pm$	1 & 0.999 & 0.034\\
Rattus norvegicus	& 3845	& 0.0039	$\pm$	0.0001 &	257	$\pm$	1 & 0.998 & 0.056\\
\hline
Adenovirus	& 517	& 0.012 $\pm$	0.001 &	83	$\pm$	3 & 0.845 & 0.46\\
Apis Mellifera	& 6958	& 0.0033	$\pm$	0.0001 &	296	$\pm$	2 & 0.995 & 0.095\\
Caenorhabditis Elegans	& 4284	& 0.0015	$\pm$	0.0001 &	647	$\pm$	8 & 0.946 & 0.58\\
Ciona Intestinalis	& 3560	& 0.002	$\pm$	0.0001 &	490	$\pm$	18 & 0.688 & 0.31\\
Danio Rerio	& 4072	& 0.0035	$\pm$	0.0001 &	288	$\pm$	2 & 0.979 & 0.12\\
Drosophila melanogaster	& 568	& 0.023	$\pm$	0.001 &	44	$\pm$	1 & 0.992 & 0.33\\
Escherichia coli	& 324	& 0.037	$\pm$	0.001 &	27	$\pm$	1 & 0.973 & 0.36\\
Oikopleura diotica	& 679	& 0.019	$\pm$	0.001 &	51	$\pm$	1 & 0.920 & 0.27\\
Saccharomices Cerevisiae	& 308	& 0.027	$\pm$	0.001 &	37	$\pm$	1 & 0.995 & 0.11\\
Tetraodon nigroviridis	& 1573	& 0.0032	$\pm$	0.0001 &	312	$\pm$	7 & 0.883 & 0.74\\
Tribolium castaneum	& 2455	& 0.0026	$\pm$	0.0001 &	388	$\pm$	3 & 0.983 & 0.46\\
\hline

\end{tabular}
}
\caption{Exponential fit of CG distributions for all organisms. 
For each organism, the maximum CG distance is shown (Max), together with the fit parameters ($d$), the goodness of fit ($r^2$)  the characteristic lengths $\lambda$ (the inverse of $d$) and the value of the normalized Chi-square cumulative function ($\chi^2$). 
All errors are expressed as $95\%$ confidence intervals, and rounded to the first significant digit. 
We observe that the Chi-square values of all higher-order organisms lie in the first decile of the cumulative function, as a demonstration of the goodness of the fit.}
\end{table}

The resulting fit parameters are shown in Table 1.
As it can be seen, the  fit is very good ($r^2>0.998$, $\chi^2 < 0.1$) for the group of organisms in Fig. \ref{HomoDNA} which showed a different CG distribution.
Moreover, for these organisms, the fit parameters are very similar, with a characteristic length ranging from $200$ to about $300$ bases (from a minimum of $214$ for \textit{Equus} to a maximum of $294$ for Mouse).
The only exception is \textit{Monodelphis}, which has a characteristic length of 452, nearly double the others.  This can be due to the fact that in this organism CpG autosomal density is very different from the other amniotes ($0.9\%$ versus $1.7-2.2\%$ \cite{Mikkelsen07}).
For the other group of organisms, once again, the situation is more heterogeneous: for example, for \textit{Apis} and \textit{Danio} the exponential function seems to fit  the empirical data well ($r^2 \geq 0.98$, $\chi^2 \simeq 0.1$) and also the characteristic lengths are comparable with the ones of the first group ($\lambda = 296$,  $\lambda=299$, respectively), while \textit{Drosophila} and \textit{Saccharomices}  have very  good fits ($r^2 \geq 0.99$, $\chi^2 \simeq 0.1$ for S. cerevisiae), but their characteristic lengths differ by an order of magnitude ($\lambda = 44, 37$ respectively).
For other organisms the exponential distribution seems unfit, even if with an heterogeneous degree of dissimilarity (see Supplementary Figure 2 for a visual inspection).

Finally we remark that some small characteristic lengths could be related to the small size of the genome of the organism (as for \textit{Adenovirus},  \textit{E. coli} and \textit{S. cerevisiae}), but this association between  parameter $d$ and genome size cannot be generalized to the other organisms.

\section{Discussion}

We have characterized the first-return time distributions of dinucleotides in DNA sequences, from a large set of organisms with different levels of complexity (from viruses to primates).

What we have found in human DNA is a striking difference for the dinucleotide CG: the inter-distance distribution of CG's has an exponential tail,  while the distributions for the other dinucleotides exhibit a power-law tail. An exponential distribution of return times is found in  a  stochastic process with a characteristic time scale, that in our case represents a characteristic distance between the dinucleotides, which is very different from a process with a power-law distribution of return times. 
This feature of the dinucleotide CG  might reflect their peculiar functional and structural role inside DNA, since CG dinucleotides are known to be the sites for DNA methylation, an epigenetic mechanism known to be involved in gene regulation and also in structural conformation of DNA chromatin.

We have extended this analysis to other 20 organisms, many of which (such as mammals) should be very similar to man in terms of DNA processing  while others (like viruses, bacteria and unicellular organisms) should be very different. Finally  for the other organisms considered in our study (such as  insects, fish and worms ), the differences are not in principle so clear.
For example, many of the chosen organisms (E. coli, C. elegans, honeybee, fruitfly, Ciona, Tribolium, Danio, Tetraodon, mouse, man) are known to have different degrees of functional similarities between each other, in terms of hortologies of the main family of enzymes governing DNA methylation processes (see \cite{Jeltsch10}), but for many of the organisms we studied such information is not actually available.

What we observe is a striking similarity between mammals (Fig. \ref{Sup}), which are known to have very similar DNA methylation processes, and also similar levels of global DNA methylation. 
The investigation of the CG cumulative distributions in these organisms showed a common exponential distribution for the long-range dinucleotide inter-distances. 
Moreover,  the characteristic lengths associated to these distributions are also consistently similar ($200<\lambda<300$), thus suggesting that the common biological mechanisms involved in CG methylation are reflected in the similar DNA structure at the scale of dinucleotide inter-distance.

The remaining organisms (Fig. \ref{Inf}) span a larger range of organismal complexity, thus the overall picture appears more heterogeneous from a biological point of view, and this is reflected in our analysis.
For example Escherichia Coli, a bacterium that does not possess similar epigenetic mechanisms  and  probably does not exploit DNA methylation processes for the same purposes as pluricellular organisms, does not present significant differences between dinucleotide distributions. The same is true  for Adenovirus. This different behavior might be justified by a different role of DNA methylation in bacteria, or by the fact that only a small portion of bacterial DNA  is affected by this process \cite{Donczew14}. Thus, our statistical approach might not be sensitive enough to highlight possible differences.
The case of fruitfly looks similar, since the difference between dinucleotide distributions is not as marked, as shown in the double logarithmic plot of Fig. \ref{EcoliDNA}.  
This is consistent with what is known about the very low levels of DNA methylation in fruitfly, and  the absence of DNMT family hortologs \cite{Jeltsch10}.

This approach has been exploited to classify organisms in phylogenetic trees \cite{Afreixo09}, but our analysis shows that it might also help  infer the presence of DNA epigenetic mechanisms in poorly characterized organisms, even if a clear association of these observations with specific biological mechanisms is yet to come.  

The exponential function used to fit of the CG inter-distance distribution at a whole-genome level seems to highlight a structural role of CGs in higher-level organisms, since it is the signature of a regular  ``marking'' along all the DNA sequence.
An observation, that in our opinion might deserve deeper investigation, is that the characteristic lengths found in the first group of organisms is comparable to the length typically associated with histones, protein complexes that play a role in chromatin modelling.
It is known that the length of a DNA sequence wrapped around histones is the size of the nucleosome plus a variable linker DNA region, summing up to about 220 bases, therefore the characteristic length of CG inter-distance could be associated with the positioning of histone positioning along the genome and possibly with three-dimensional  structure of DNA.

\section*{Authors' contribution}
\noindent GP generated the data and analyzed the data. GC designed the analysis and analyzed the data. BM, ML, MDE and GCC helped interpreting the results and revised the paper. DR designed the analysis, analyzed the data and wrote the paper.
The Authors declare no financial or other competing interests.

\section*{Funding}
\noindent DR, GC, BM and ML were partially funded by the University of Bologna grant "FARB Linea 1" 2013-2016. DR and GCC were also partially funded by the MIUR Flagship InterOmics grant (PB05).

\bibliographystyle{plain}

\end{document}